\documentstyle[prb,aps,epsfig]{revtex}

\begin{document}

\twocolumn  

\noindent {\bf Coulomb interaction and instability of CE-type structure in
half-doped manganites}

In their Letter,[1] den Brink, Khaliullin, and Khomskii proposed
theoretically that the one-dimensional ferromagnetic zigzag chains in CE
phase in half-doped manganites play an essential role in forming the orbital
ordering, and, more surprisingly, the on-site Coulomb interaction U between
electrons with different orbitals leads to experimentally observed charge
ordering. In this Comment, I point out that the strong U will destroy the
stability of CE-type phase, which is stable in a very narrow regime in the
parameter space for electronic model.

The same notations as in Ref. [1] are employed here. In Ref. [1], the case
of $U=0$ was analyzed and CE phase was found to be stable when the
superexchange coupling $J>0.1524t$ (t is the hopping integral). In the CE
phase the one-dimensional zigzag chain consists of two types of lattice
sites, the bridge and corner sites. The bridge site is occupied singly by
electrons with certain orbitals while the corner sites can be occupied
doubly. Without U, the charge distribution along the zigzag chain is
uniform, i.e., 1/2. Finite U will push the charge away from the corner sites
to the bridge sites to form charge ordering. The theory explains the
patterns of three orderings almost perfectly. We observe that the U will
also increase the energy of the CE phase. The rod-type (i.e., C-type) phase
consists of one-dimensional ferromagnetic straight chains. The chains are
occupied at most singly on each site. Thus the U does not change the energy
of the C-type phase. This fact leads to the instability of CE phase with
respect to C-type phase for finite U. We evaluate the electronic energy for
CE phase by using exact diagonalization method. The energies from the
magnetic coupling are the same for both C-type and CE phase. At $U=0$, $%
E_{kin}^{CE}=-0.6953t$ for CE-type phase and $E_{kin}^{C}=-0.6366t$ for
C-type phase. However, at $U=+\infty $, $E_{kin}^{CE}=-0.61t$ and the energy
of C-type phase keeps unchanged. In other words, the CE phase is unstable
with respect to C-type phase when electron correlation becomes stronger.
Numerical results for energies of CE phase via the U is shown in Fig. 1(a).
The calculation is performed on 8-, 12- and 16-site chains, respectively.
The size-effect is very small. The main reason is that the half-doped zigzag
chain is a band insulator. This is also seen from the calculation of $%
E_{kin}^{CE}$ for a finite chain when $U=0$. Its value for 16 sites is
-0.695328t, which is equal to the value of an infinite chain up to the order
of 10$^{-6}$.[2] At about $U\approx 5t$, the CE-type phase has a higher
energy than C-type phase. To establish a comprehensive phase diagram, we
have to consider other possible phases. Except for conventional anti- and
ferromagnetic phases, a novel type of phase is taken into account. The phase
consists of two-site ferromagnetic valence bonds (VB), and the bonds couple
antiferromagnetically. Electrons are localized within the bonds. The
electronic part of the state in a valence bond is $\frac{1}{\sqrt{2}}%
(c_{i,\alpha ,\sigma }^{\dagger }+c_{j,\alpha ,\sigma }^{\dagger })\left|
0\right\rangle $, and the local spins are parallel to spin of electron. The
energy per site is $E_{kin}^{VB}+E_{mag}^{VB}=-0.5t-J$, which is independent
of U (we take the magnetic energy as zero for CE phase). The ferromagnetic
phase has energy $-t+2J$ for three dimension and $-0.919t+2J$. Thus when $U=0
$ and $0.152t\leq J\leq 0.195t$ for three dimension and $0.117t\leq J\leq
0.195t$ for two dimension, CE phase has a lower energy. The phase diagrams
for $U=0$ and $U=+\infty $ as shown in Fig.1(b) and (c) are evaluated on a 4$%
\times $4 lattice\ with a periodic boundary condition by means of
combination of mean field theory and exact diagonalization. We assume the
magnetic structure and claculate the energy by exact diagonalization. FM and
C-AF are assumed to contain two sublattices and evolve into G-AF when spins
on two sublattices are not parallel for large J. Thus, the CE phase does not
appear in the ground state for the model in Ref. [1] in the case of strong
correlation.

In doped manganites, U is usually estimated to be much larger than t. In
Ref. [1], $U\approx 10t$. The on-site U alone can produce the experimentally
observed pattern of charge ordering, but cannot stabilize the CE phase. To
solve this issue finally, we have to take into account other interactions,
such as the long-range Coulomb interaction, Jahn-Teller distortion, and
physics of topological berry phase. For example, the effect of finite large J%
$_{H}$ leads to an attractive particle-hole interaction, which favors to
stabilize the charge ordering.[3]

This work was supported by RGC grant of Hong Kong.

\begin{figure}[tbp]
\centerline{\epsfig{file=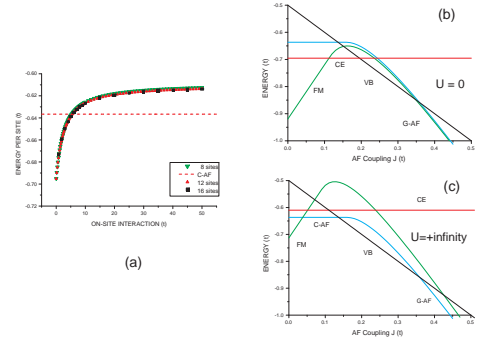, width=7.5cm}}
\caption{(a) The electronic energy of CE phase per site via the
on-site interaction; (b) and (c) the phase diagrams on a square lattice for $%
U=0$ and $+\infty ,$ respectively .}
\end{figure}

\noindent Shun-Qing Shen

Department of Physics, The University of Hong Kong, Hong Kong and Institute
of Physics, Chinese Academy of Science, P. O. Box 603, Beijing 100080, China


\begin{references}
\bibitem{note}  The finite-size effect for $E_{kin}^{C}$ is relatively
large. $E_{kin}^{C}(12)=-0.6220t$ while $E_{kin}^{C}(+\infty )=-0.6366t$. So
the calculated value $U_{c}$ from Fig. 1(a) should be more reasonable than
that from 12 sites.
\end{references}
\end{document}